\documentclass[10pt]{article}
\usepackage[utf8]{inputenc}
\usepackage[T1]{fontenc}   
\usepackage{amssymb,amsmath,bbm}
\usepackage{enumerate}
\usepackage{resizegather}
\usepackage{xcolor}
\usepackage{graphicx}
\usepackage{amsthm}
\usepackage{comment}
\usepackage{hyperref}
\usepackage{caption}
\usepackage{subcaption}
\usepackage{float}
\usepackage{csvsimple}
\usepackage{longtable}
\usepackage{natbib}

\newtheorem{oss}{Remark}

\def\P{\mathbb P}
\def\E{\mathbb E}

\begin{document}
\title{How network properties and epidemic parameters influence 
	stochastic SIR dynamics 
	on scale-free random networks}

\maketitle
\begin{center}
	
\author{Sara Sottile{$^{\dagger}$}, Ozan Kahramano{\u g}ullar{\i}{$^{\dagger,}\!$}
	\footnote{Corresponding author: \texttt{ozan.kah@gmail.com}}\,, Mattia Sensi{$^{\dagger}$}\\[1em]
	$^\dagger$\normalsize  Department of Mathematics, University of Trento,\\
	\normalsize Via Sommarive 14, 38123 Povo - Trento, Italy}
\end{center}

\begin{abstract}
With the premise that social interactions are described by power-law distributions, we study a SIR stochastic dynamic on a static scale-free random network generated via configuration model. We verify our model with respect to deterministic considerations and provide a theoretical result on the probability of the extinction of the disease.
Based on this calibration, we explore the variability in disease spread by stochastic simulations. In particular, we demonstrate how important epidemic indices change as a function of the contagiousness  of the disease and the connectivity of the network. Our results quantify the role of starting node degree in determining these indices, commonly used to describe epidemic spread.
\end{abstract}


\section{Introduction}\label{Intro}
Real systems of interactions between humans are commonly studied as interaction networks. In this setting,  an edge in a network represents a possible interaction between the connected nodes.  The mechanism of disease transmission upon contact provides a strong case for the use of these networks in modelling epidemic. Spreading processes of infectious diseases take place over these networks, thus their structures can be essential for understanding disease transmission. In particular, knowledge on the interplay between network structure and transmission parameters can improve predictions and prevention and control strategies.

A key factor in epidemic spread is the population network structure where a disease may spread. However, a complete picture, in principle, may require the knowledge of every individual in a population and its relationships, for example, as in contact tracing \cite{contact_tracing}. Networks used in different fields have some common geometric characteristics regarding the distribution of the nodes and edges, for example, a few nodes that may act as hubs and the vast majority of nodes with a few neighbours \cite{intro}. A classification in this context is given by network generation models, used as generators of synthetic networks, with controlled topological properties. Several types of networks have been proposed over time \cite{Keeling_intro,Moreno}. A simple model was built by Erd{\" o}s and R{\' e}nyi \cite{Erdos} where nodes are connected according to a uniform probability without any preference. However, some structural properties observed in real-world world networks cannot be reproduced by this model as empirically verified by Newman \cite{NewmanPark}. It has been observed in real networks that the degree distribution of individuals is far from homogeneous and only a few individuals have several connections and the majority have a few connections \cite{albertetall1,albertetall2}. Barab{\'a}si and Albert proposed a model to generate \textit{scale-free networks} \cite{Barabasi} with a connection mechanism that mimics the natural formation of social contacts. However, their algorithm cannot control the value of the exponent of the power-law distribution \cite{Keeling_intro}. The \textit{configuration model} can overcome this issue, generating a network with a given degree sequence. This model is used, for example, in social dynamics as it captures connectivity features of this class of networks \cite{moreno2}. The study of how epidemic evolve on networks has been tackled with various approaches, theoretical and computational. Theoretical results on this topic can be found, for example, in the monographs \cite{Estrada, Miller}. In \cite{lopez2016stochastic}, the author derives results on a small heterogeneous population before validating them via numerical simulations. In \cite{strona2018intrinsic}, the authors explore the interplay between network properties and disease characteristics on the spread of the latter. Many other articles are devoted to the study of the impact of the network structure on the evolution of an epidemic \cite{ball2010analysis, draief2006thresholds, ganesh2005effect, zhang2013stochastic}. Another interesting feature, studied in \cite{rocha2011simulated}, is the implementation of dynamic contacts, meaning a network in which nodes may delete and create edges in time.

To this end, in this paper we investigate how the interplay between the connectivity of different configuration model networks and the contagiousness of the disease affect the magnitude of the epidemic. 
Following a  summary of the relevant mathematical background, 
we derive a lower bound and a closed formula for the probability of extinction of the disease in a network, as a function of all the parameters involved and the degree of the initially infected node. 
We introduce a stochastic model on scale-free random network, which we use to run simulations on different instantiations of the model.
The simulations, carried out via a specialized modification of the Gillespie algorithm, validate our theoretical results regarding the probability of extinction of the disease.

We provide a  thorough analysis via simulations that highlights the influence of network connectivity and infectiousness on three key epidemic indices: the maximum number of simultaneously infected individuals, the total number of eventually infected individuals, and the duration of the epidemic. In particular, we focus on the role of the position of the nodes from which the disease starts spreading as well as the number of initially infected nodes, exploring simulations for different instantiations  of the model parameters. 
We compare the evolution of the epidemic, measured by the aforementioned indices, in four different cases for the initial infected nodes, categorized by their position in the respective network: hub (degree in the tail of the distribution), mean degree, peripheral (low degree), and randomly chosen. The stochastic simulations with our model, in agreement with the deterministic lower bound for the probability of extinction, provide a quantification of the role of model  parameters in the spread of the epidemic on scale-free networks. Overall, our results demonstrate how the combination of deterministic considerations and stochastic simulations can be used to provide insights on epidemic spread.

\section{Power-law networks}\label{pwl}
In this Section, we describe scale-free networks, which can be used to mathematically evaluate a given model or describe a real network structure, mimicking social contacts between individuals. We then describe the classic algorithm used to generate networks with a certain degree distribution,
that is, \emph{configuration model} (CM).

\subsection{Scale-free networks}\label{Scale}
Many real-world graphs follow power law degree distributions, that is, degree distributions with the probability of a node to have $k$ direct neighbours given by $p_k \sim k^{-\alpha}$. These kind of networks are also called \textit{scale-free random graphs}. The preferential attachment model of Barabási and Albert \cite{Barabasi} produces a degree distribution with $\alpha = 3$. However, there are many examples in which $\alpha \in (2,3]$ (see e.g. \cite[Sec. 1.4]{Durrett}). To explain the scale-free property, Barabási defined the power-law distribution both with a discrete and a continuum formalism; we refer to \cite{Barabasi_book} for a more in-depth introduction to this topic, and recall here the definitions and results we need for our analysis.

In a scale-free network, the probability of having $i$ neighbours is
\begin{gather*}
p_k = C k^{-\alpha},
\end{gather*}
where $C$ is the normalization constant such that 
\begin{gather*}
\sum_{k_{\min}}^{\infty} p_k = 1,
\end{gather*}
where $k_{\min}$ is the smallest degree we allow a node to have.

The main difference between a random and a scale-free network comes in the tail of the degree distribution, representing the high-$k$ region of $p_k$: high-degree nodes, called \textit{hubs}, are naturally present in scale-free networks, contrary to random networks. Since all real networks are finite, we may expect that nodes assume a maximum degree, $k_{\max}$, called the \textit{natural cut-off of the degree distribution $p_k$} (see \cite{cutoff}). This quantity represents the expected size of the largest hub in a network. To calculate $k_{\max}$, we assume that in a network $\mathcal{G}(N,V)$ of $N$ nodes we expect at most one node in the $(k_{\max}, \infty)$ regime. We proceed as in \cite{Barabasi_book}, and set ourselves in the continuum formalism. In this case, the probability to observe a node whose degree exceeds $k_{\max}$ is
\begin{gather*}
\int_{k_{\max}}^\infty p(k) \textnormal{d}k \sim \dfrac{1}{N}.
\end{gather*}
Since the degree distribution has the form of
\begin{gather*}
p(k)  \sim  k^{-\alpha},
\end{gather*}
the natural cut-off follows
\begin{gather}
k_{\max} = \lfloor k_{\min} N^{\frac{1}{\alpha-1}} \rfloor .
\label{eqn:kmax}
\end{gather}
Hence, the highest attainable degree is directly proportional to a power of $N$ between $\frac{1}{2}$ (corresponding to $\alpha=3$) and $1$ ($\alpha=2$). Once we have obtained this values for $k_{\max}$, we can compute again the normalization constant $C$ such that
\begin{gather*}
\sum_{k_{\min}}^{k_{\max}} p_k = 1.
\end{gather*}

\subsection{Configuration model}\label{Config}

In the case of large networks, the adjacency matrix is typically not available. However, in the case of real networks of which we know the degree sequence, we can generate a graph with precisely the same degrees. Given the number of nodes $N$ and the sequence of degrees $\{k_i\}_{1\leq i \leq N}$ of length $N$ (we omit the subscript from now on, for ease of notation), the aim is to construct an undirected graph with $N$ vertices in which the $i$-th vertex has precisely degree $k_i$. We denote such graphs with $\mathcal{G}(N,\{k_i\})$. Given a degree distribution obtained from observing a stochastic network, the algorithm used to fit this distribution is called \textit{configuration model} (CM).
To construct the network, we start by assigning to each vertex $i$ in the set of nodes a random degree $k_i$, drawn from the chosen probability distribution $p(k)$. Clearly, $k_{\max} \leq N-1$ since no vertex can have a degree larger than $N-1$. The degrees of the vertices are represented as half-links or stubs, thus we impose the constraint that the sum $\sum_{i=1}^N k_i$ must be even, that is 
\begin{gather*}
\sum_{i=1}^N k_i = 2m,
\end{gather*}
for some $m \in \mathbb{N}$.

First, two stubs are connected to form an edge. After that, another pair of stubs are chosen from the remaining $2m - 2$ stubs and connected, respecting the preassigned degrees. The network is completed by repeating this procedure until the stubs run out. The result of this construction is a random network whose degrees are distributed according to $p_k$ \cite{Newman}. If $L$ denotes the numbers of degrees assumed in the network and $N_1$, $N_2$, $\dots$, $N_L$ the number of nodes of each degree, the average degree in the network is given by \cite{Miller}
\begin{gather*}
\langle k \rangle = \dfrac{1}{N}\sum_{i = 1}^L N_i \Tilde{k}_i,
\end{gather*}
where $\Tilde{k}_i$ are the different degrees of the nodes in the graph, listed from the sequence $\{k_i\}$. This formula is equivalent to $\langle k \rangle = \sum_k k p_k$, for this specific realization. Note that this linkage procedure does \textit{not} exclude self-loops or multiple edges, but their expected number is bounded (see e.g. \cite[Prop. 7.1]{Hofstad}). When the size of the graph $N \to +\infty$ with a fixed degree distribution, self-loops and multiple edges become less and less apparent in the global dynamics (see e.g \cite[Th. 3.1.2]{Durrett}).

Let us remark the fact that a \textit{giant component}, a connected component of the network that contains a significant proportion of the entire nodes, exists if and only if
\begin{gather}\label{gian}
\langle k^2 \rangle - 2 \langle k \rangle > 0.
\end{gather}
In power-law distributions, the second moment $\langle k^2 \rangle$ diverges for any chosen exponent $\alpha \in (2,3)$ whereas the first moment $\langle k \rangle$ is finite.
This means that (\ref{gian}) is always satisfied for any configuration model with a power-law tail to its degree distribution as long as $2 < \alpha < 3$. Hence, there will always be a giant component. However, the giant component is not necessarily connected to the whole
network, and the remaining nodes may form smaller components \cite{Newman}.

\section{Lower bound for the probability of extinction}\label{theoretical}
We work on an SIR epidemiological dynamics model on a network built with the CM model algorithm, described in Section \ref{Config}. 
In this network, we introduce a number of infected individuals in an otherwise fully susceptible population; 
the exact number of infected individuals is specified each time it changes. 
Later, in Section \ref{Basic}, we study how the introduction of a different number of initially infected individuals affects the evolution of the epidemic. 
In this section, we consider an epidemic which starts with only one infected individual, and we derive a deterministic formula for the \textit{probability of extinction}, that is, the probability that one infected individual in the network does not cause a major outbreak. We use this probability as a benchmark to compare with our simulations. As usual, we assume that an individual remains infected for an exponentially distributed time of parameter $\gamma$. During its infectious period, an individual infects each of its neighbours (independently of the others) according to a Poisson process of parameter $\beta$. 
Note that modellers often assume that $\beta$ does not depend on the number of contacts; however, we notice that in most epidemic models (and data that are collected) the infection probability decreases with the number of contacts. The \emph{basic reproduction number} $R_0$, in this setting, is given in \cite{LibroPugliese}, in which the expression is computed as
\begin{gather*}
R_0 = \frac{\beta}{\gamma}\langle k \rangle (1 + C_v^2),
\end{gather*}
where $C_v$ is the coefficient of variation, defined as $C_v =  \frac{\langle k^2 \rangle}{\langle k \rangle}$. Using this definition, the expression of $R_0$ can be written also as
\begin{gather}
R_0=\frac{\beta}{\gamma}\frac{\langle k^2 \rangle}{\langle k \rangle}.
\label{eqn:rnod}
\end{gather}

We focus on this slightly unrealistic, but analytically tractable assumption. We approximate the initial phase of an epidemic by a branching process where all contacted individuals are susceptible. This is a standard approximation in epidemic models; thus, cliques and triangles are neglected, and the population is assumed to be large enough. First, we compute the probability that an infected with $i$ neighbours infects $k$ of them. 
We start by considering one of them, conditioning on the length of the infectious period:
\begin{gather*}
\P( \mbox{a contact is not infected}) = \gamma \int_0^\infty e^{-\beta t} e^{-\gamma t}\, \textnormal{d}t = \frac{\gamma}{\beta + \gamma  } = \frac{1}{R +1 },
\end{gather*}
where $R:= \beta/\gamma$.\\
We can not now use the binomial distribution, because infections of different contacts are not independent, but correlated by the length of the infectious period. This is due to the fact that if the infectious period is short, it is likely that no neighbour will be infected, while if it is long, most of them will. Indeed, if $Q$ is the number of infected neighbours,
\begin{gather*}
\P(Q = 0)  = \gamma \int_0^\infty (e^{-\beta t} )^i e^{-\gamma t}\,\textnormal{d}t  = \frac{\gamma}{ \beta i + \gamma  } = \frac{1}{i R +1 }.
\end{gather*} 
The other expressions are more complicated:
\begin{gather}\label{PQk}
\P(Q = k)  = \gamma  \int_0^\infty {i \choose k} (1 -e^{-\beta t})^k (e^{-\beta t} )^{i-k} e^{-\gamma t}\, \textnormal{d}t,
\end{gather}
clearly, $  \displaystyle \sum_{k=0}^{i } \P(Q = k) = 1 $, indeed
\begin{align*}
\sum_{k=0}^{i} \P(Q= k) =&  \sum_{k=0}^{i} \gamma\int_{0}^{\infty}\binom{i}{k}(1-e^{-\beta t})^k (e^{-\beta t})^{i-k} e^{-\gamma t} \textnormal{d}t \\
=& \gamma \int_{0}^{\infty}e^{-\gamma t} \bigg(\sum_{k=0}^{i} \binom{i}{k}(1-e^{-\beta t})^k (e^{-\beta t})^{i-k} \bigg) \textnormal{d}t.
\end{align*}
Note that $\displaystyle \sum_{k=0}^{i} \binom{i}{k}(1-e^{-\beta t})^k (e^{-\beta t})^{i-k} = (1- e^{-\beta t}+e^{-\beta t})^i=1,$ thus we obtain
\begin{equation*}
\sum_{k=0}^{i} \P(Q= k) = \gamma \int_{0}^{\infty}  e^{-\gamma t} \textnormal{d}t 
= \gamma \lim_{a \to \infty} \bigg(-\frac{e^{-\gamma a}}{\gamma} + \frac{1}{\gamma}\bigg)  = 1.
\end{equation*}
We can clarify the expression in \eqref{PQk} as
\begin{align*} \P(Q = k)  =& \gamma  \int_0^\infty {i \choose k} (1 -e^{-\beta t})^k (e^{-\beta t} )^{i-k} e^{-\gamma t}\, \textnormal{d}t  = (\mbox{integr. by parts})\\
=& \frac{\gamma \beta}{\gamma + \beta(i-k) } \frac{i !}{ (i-k)! (k-1)!} (1 -e^{-\beta t})^{k-1} (e^{-\beta t} )^{i-k+1} e^{-\gamma t}\, \\
=& \frac{\beta (i-k+1)}{\gamma + \beta(i-k) } \P(Q = k-1) .
\end{align*}
Then, by induction, we obtain
\begin{gather}
\label{Q_k}
\P(Q = k) = 
\frac{R^k  \prod_{j=1}^k  (i-j+1)}{\prod_{j=0}^k (1+R(i-j))  } =: r_k.
\end{gather}
Although the  infections of different contacts are not independent, the expected number of contacts infected, $\E(Q)$ can be easily obtained as 
\begin{gather*}
\frac{i R}{R +1 }\,
\end{gather*}
because the expected value of a finite sum is the sum of the expected values. We need these computations in order to estimate the probability of extinction of the branching process, to which we refer as a \emph{minor epidemic}. Indeed, we have a multitype branching process (types are the number of neighbours) and its probability of extinction is given by the smallest positive solution of $ s = f(s)$ where $s=(s_1,\ldots, s_N)$ (if $N$ is the number of types) and
\begin{gather*}
f_i(s) = \sum_{\mathbf{k}=(k_1,\ldots,k_N)}^{ } p_i(\mathbf{k}) s_1^{k_1} \cdots s_N^{k_N},
\end{gather*}
with
\begin{gather*}
p_i(\mathbf{k}) = \P(\mbox{an infected of type }i\mbox{ infects }k_1\mbox{ of type }1,\ldots,\ k_N\mbox{ of type }N).
\end{gather*}

To be precise, once we found the smallest solution $s^*$, $s_i^* = f_i(s^*)$ represents the extinction probability starting with one individual of type $i$. For the derivation of these formulae, we refer to \cite{extinction1,extinction2}.
Note that $f_i(1,\ldots, 1) = 1$; hence $(1,\ldots, 1)$ is always a solution of $f(s)=s$. If we find a smaller solution, then the probability of extinction is smaller than 1; otherwise, the probability of extinction is 1.
We then need to compute $p_i(\mathbf{k})$. An individual with $i$ neighbours will have $m_1$ of type 1, \ldots, $m_N$ of type $N$ (where $m_1 + \cdots +m_N = i$) with probability
\begin{gather*}
\frac{i!}{m_1! \cdots m_N! } q_1^{m_1}\cdots q_N^{m_N},
\end{gather*}
where 
\begin{gather*}q_j  = \frac{jp_j}{\sum_{k=1}^{N } k q_k } \qquad\mbox{(size-biased probabilities)}.
\end{gather*}
We could then compute the probability of infecting $k_1$ out of $m_1$, $k_2$ out of $m_2$, \ldots, $k_n$ out of $m_n$ and sum over all possible combinations. We do not need to go through all that because the probability of infecting one neighbour is independent of its properties (i.e. of its degree). Hence,
if $k = k_1 + \ldots + k_N$, 
\begin{gather*}
p_i(\mathbf{k}) = \P(Q = k) \frac{k!}{k_1! \cdots k_N! } q_1^{k_1}\cdots q_N^{k_N}.
\end{gather*}
We compute the probability of infecting $k$ neighbours, and then that of the $k$ infected neighbours, $k_1$ were of type 1,\ldots $k_N$ of type $N$. We can go a further step in the computation of $f_i(s)$; indeed
\begin{align}
f_i(s) =& \sum_{\mathbf{k}=(k_1,\ldots,k_N)}^{ }  \P(Q = k) \frac{k!}{k_1! \cdots k_N! } q_1^{k_1}\cdots q_N^{k_N}s_1^{k_1} \cdots s_N^{k_N} \nonumber
\\
=& \sum_{k=0}^i  \P(Q = k) \sum_{\mathbf{k}:\ k_1+\cdots+k_N=k }\frac{k!}{k_1! \cdots k_N! } (q_1s_1)^{k_1}\cdots (q_N s_N) ^{k_N} \nonumber\\
=& \sum_{k=0}^i  \P(Q = k) (q_1 s_1 + \cdots + q_n s_n)^k \nonumber\\
=& \sum_{k=0}^{i } \left(R(q_1 s_1 + \cdots + q_n s_n)\right)^k \frac{  \prod_{j=1}^k  (i-j+1)}{\prod_{j=0}^k (1+R(i-j))  }. \label{Formula}
\end{align}

One could go beyond this, but the formulae would become increasingly cumbersome. In order to compute the relevant solution numerically, one can start with a vector $\mathbf{s^0} \le s^*$ (for instance $\mathbf{s^0} = 0$) and then compute $\mathbf{s^n} = f(\mathbf{s^{n-1}})$. Iterating this, we converge to the required fixed point.
\begin{oss}
	However, one may notice one property; obviously $s_i^* = f_i(s^*) > f_i(0) = \frac{1}{1 + R i }$. For instance, if $R = 0.002$, as in the first line of the example, and $i \le 50$, we get 
	$s_i^* > \frac{1}{1.1 }\approx 0.91$, thus we immediately see that the extinction probability is very high. Hence, a lower bound for the probability of extinction is given by
	\begin{gather}\label{eqn:loweee}
	l:=\frac{1}{1 + R i }.
	\end{gather}
	We recall that $R=\beta/\gamma$ and $i$ is the degree of the initially infected node.
\end{oss}
In order to see if the probability of extinction is 1 or lower, we can resort to $R_0$, the spectral radius of the $N \times N$ matrix $M$, whose elements $m_{ji}$ are the expected number of infected of type $j$ generated by an infected of type $i$. This is very easy to compute, conditioning on the number of neighbours:
\begin{align*} m_{ji} =& \E\left(\E(\mbox{infected of type }j | k \mbox{ neighbours of type }j)\right) \\=& \E\bigg( \frac{kR}{ R+1} \bigg) =\frac{i q_j R}{ R+1}  = \frac{R}{R+1 } \frac{i j p_j}{\sum_{k=1}^{ N} kp_k}.  \end{align*}
$M$ is a matrix of rank 1; hence its spectral radius is easy to compute. From
$ Mv = \rho v$, we obtain
\begin{gather*}
\sum_{i=1}^{n }m_{ji} v_i =  \frac{R}{R+1 } \frac{j p_j}{\sum_{k=1}^{ N} kp_k} \sum_{i=1}^{N } i v_i = \rho v_j.
\end{gather*}
Since exists $C > 0 $ such that $v_j = C j p_j$, thus
\begin{gather*}
C \frac{R}{R+1 } \frac{j p_j}{\sum_{k=1}^{ N} kp_k} \sum_{i=1}^{N } i^2 p_i  = C \rho j p_j,
\end{gather*}
which implies
\begin{gather*}\rho = \frac{R}{R+1 } \frac{\displaystyle\sum_{i=1}^{N } i^2 p_i}{\displaystyle\sum_{k=1}^{ N} kp_k}.
\end{gather*}
This is exactly the expression in (\ref{eqn:rnod}).

\section{The stochastic model}\label{Basic}

We build a network following the CM algorithm\footnote{All the codes and additional data are available on \url{https://github.com/SaraSottile/StochasticSIRnetwork}.}
\footnote{Animations of sample simulations with different rates are available at
	\href{https://www.youtube.com/playlist?list=PLdDHYeVsbaLUY7-9gt9F01JEgIFm8D09m}{https://www.youtube.com/playlist?list=PLdDHYeVsbaLUY7-9gt9F01JEgIFm8D09m}}, 
choosing $k_{\min} =1$, and fixing $N=10000$; $k_{\max} $ then varies with $\alpha$ as described by \begin{equation*}
k_{\max} = \lfloor k_{\min} N^{\frac{1}{\alpha-1}} \rfloor .
\end{equation*}

\begin{figure}[H]
	\centering
	\includegraphics[width=0.85\textwidth]{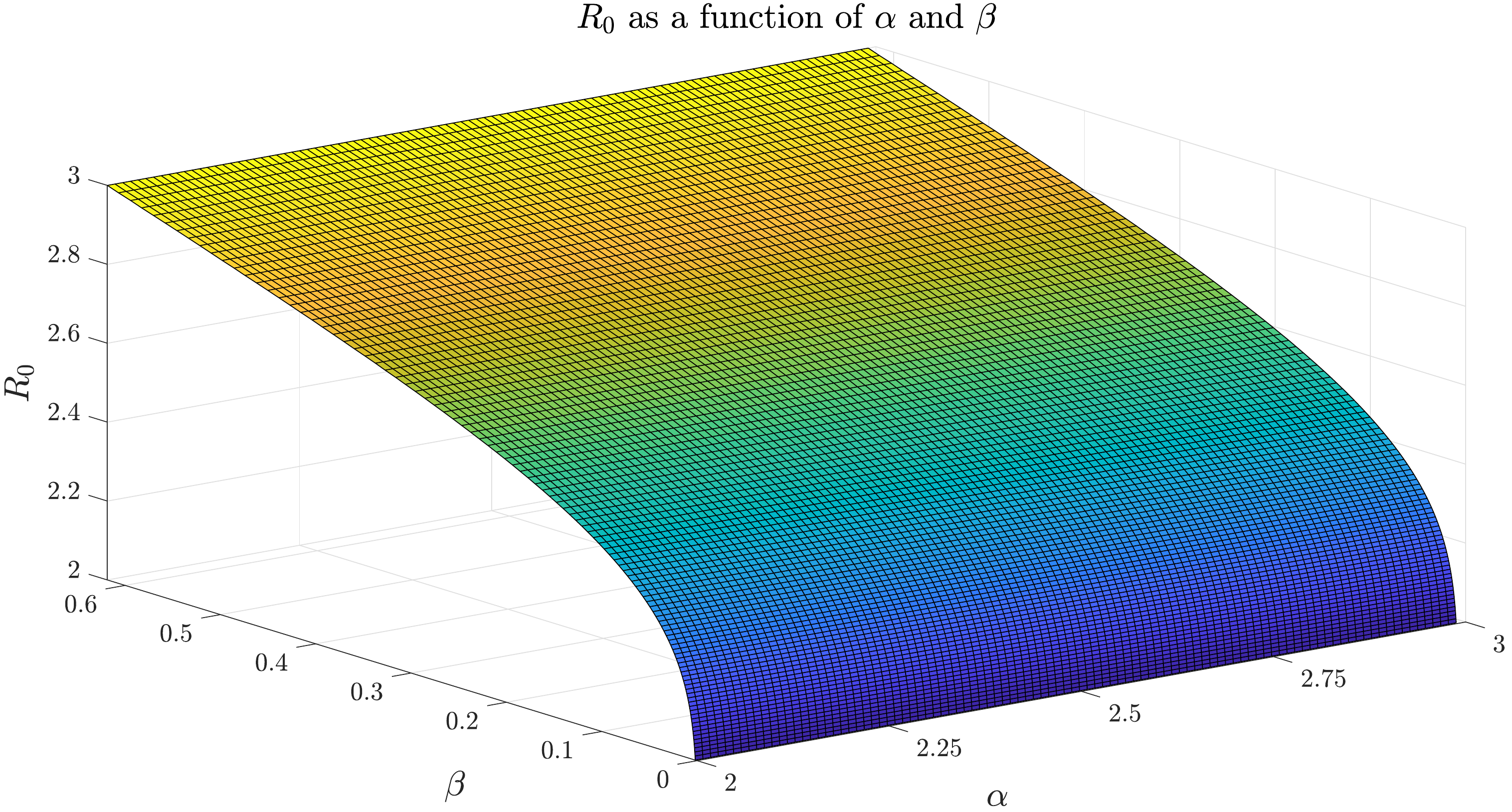}    \caption{Plot of $R_0$ for the intervals of $\alpha$ and $\beta$ we are interested in.}
	\label{fig:Rzero}
\end{figure}

We focus on the interval $2\leq R_0 \leq 3$; its value varies with the parameters $\alpha \in [2,3)$ and $\beta$ as shown in Figure \ref{fig:Rzero}, following \eqref{eqn:rnod}.

We simulated the spread of the epidemic on a network with the following values for the parameters:
\begin{itemize}
	\item $\alpha \in \{2.0, 2.11, 2.22, 2.33, 2.44, 2.55, 2.66, 2.77, 2.88, 2.99\}$;
	\item $\beta \in \{0.002, 0.071, 0.139, 0.208, 0.276, 0.345, 0.413, 0.482, 0.55, 0.619\}$;
	\item $\gamma = 1$, without loss of generality as it scales with the chosen time unit.
\end{itemize}

The simulation algorithm implements a specialised version of the Gillespie algorithm \cite{Gil77} that makes use of the network structure for enhancing the simulation efficiency. This refinement provides an equivalent algorithm for interactions on a network as in our model, while drastically reducing the simulation times in comparison to an implementation that accommodates the standard Gillespie algorithm. The main difference in our implementation is in the computation of the second order reaction propensities: at any time, an infection can occur as an instantiation of a second order reaction. The propensity of an infection is proportional with the number of possible infection events. However, on a network, the number of possible infection events is given by the the number of edges that connect the susceptible and infected nodes in the network. By exploiting this fact, our algorithm computes the propensity of the second order reactions as the product of the number of these edges and the reaction 
rate constant $\beta$. The propensities of the first order reactions are computed as usual as the product of the number of infected nodes and the rate constant $\gamma$.    

For each pair or parameters listed above, we simulated the epidemic $100$ times. In each simulation, we recorded the maximum number of \textit{simultaneously infected} nodes and the number of \textit{eventually infected}, that is, the total number of recovered at the end of the epidemic with $I=0$ in the SIR model. For each pair of parameters, we computed the average of these values, depicted below.

\subsection{Deterministic vs. stochastic ``hub case''}
In this section, we verify our stochastic model with respect to the deterministic results  
in Section \ref{theoretical} for the case in which the infection starts from a hub of the network. 
The first infected node is chosen as the node with the maximum degree 
in the network whereby the degree is inversely proportional to  $\alpha$: 
this follows from expression \eqref{eqn:kmax}.

We first evaluated the probability of extinction in 100 simulations 
for each pair of parameters $\alpha$ and $\beta$ and then we considered 
different threshold values to have a \textit{minor epidemic} where the 
number of eventually infected remains small; the threshold values 
considered in this study are 1, 2, 3 and 4.
We then computed the difference between this data and the deterministic 
probability of extinction. Finally, we computed the mean and the standard 
deviations in order to choose the ``best threshold value'' for the comparison. 
\begin{table}[H]
	\caption{Mean and Standard deviation of the difference between the stochastic simulations and deterministic  results.}
	\label{tab:table}
	\begin{tabular}{|c|c|c|c|c|}
		\hline
		& \textbf{Threshold 1} & \textbf{Threshold 2 }& \textbf{Threshold 3} & \textbf{Threshold 4 }\\ \hline
		\textbf{ Mean} & -0,06 & -0,02 & 0,00 & 0,02 \\ \hline
		\textbf{Sd} & 0,09 & 0,05 & 0,04 & 0,05 \\ \hline
	\end{tabular}
\end{table}
From Table \ref{tab:table}, we observe the best agreement when the threshold is set to 3. After choosing the threshold to define the minor epidemic, we compared  the deterministic and simulation results to verify our model. This comparison is depicted in Figure \ref{fig:comparison}. The simulation results are in good agreement with the deterministic results, however they do not coincide exactly. This can be explained by the fact that \emph{theoretically} we consider a network with an infinite number of nodes, but \emph{practically} we use a finite network with a large number of nodes. In Figure \ref{fig:comparison} we compare, for the probability of extinction of the epidemic, the deterministic value (blue) obtained in Section \ref{theoretical} with the stochastic realizations (red) with simulations. For each ($\alpha$, $\beta$) couple, we performed 100 simulations.
We observe that for the smallest value of $\beta$, the probability of extinction seems to be almost independent of the connectivity of the network. Moreover, when $\beta =0.002$, even in the cases in which the epidemic does not immediately die out, the number of eventually infected individuals in each of the $100$ iterations do not exceed $15$, which can be seen in detail in the supplementary material. However, when the transmission rate increases, the probability of extinction increases for greater values of $\alpha$. This behaviour can be noted also in Fig. \ref{fig:het2,2} (``Hub node'').

\begin{figure}[H]
	\centering
	\begin{subfigure}[b]{0.49\textwidth}
		\centering
		\includegraphics[width=1\textwidth]{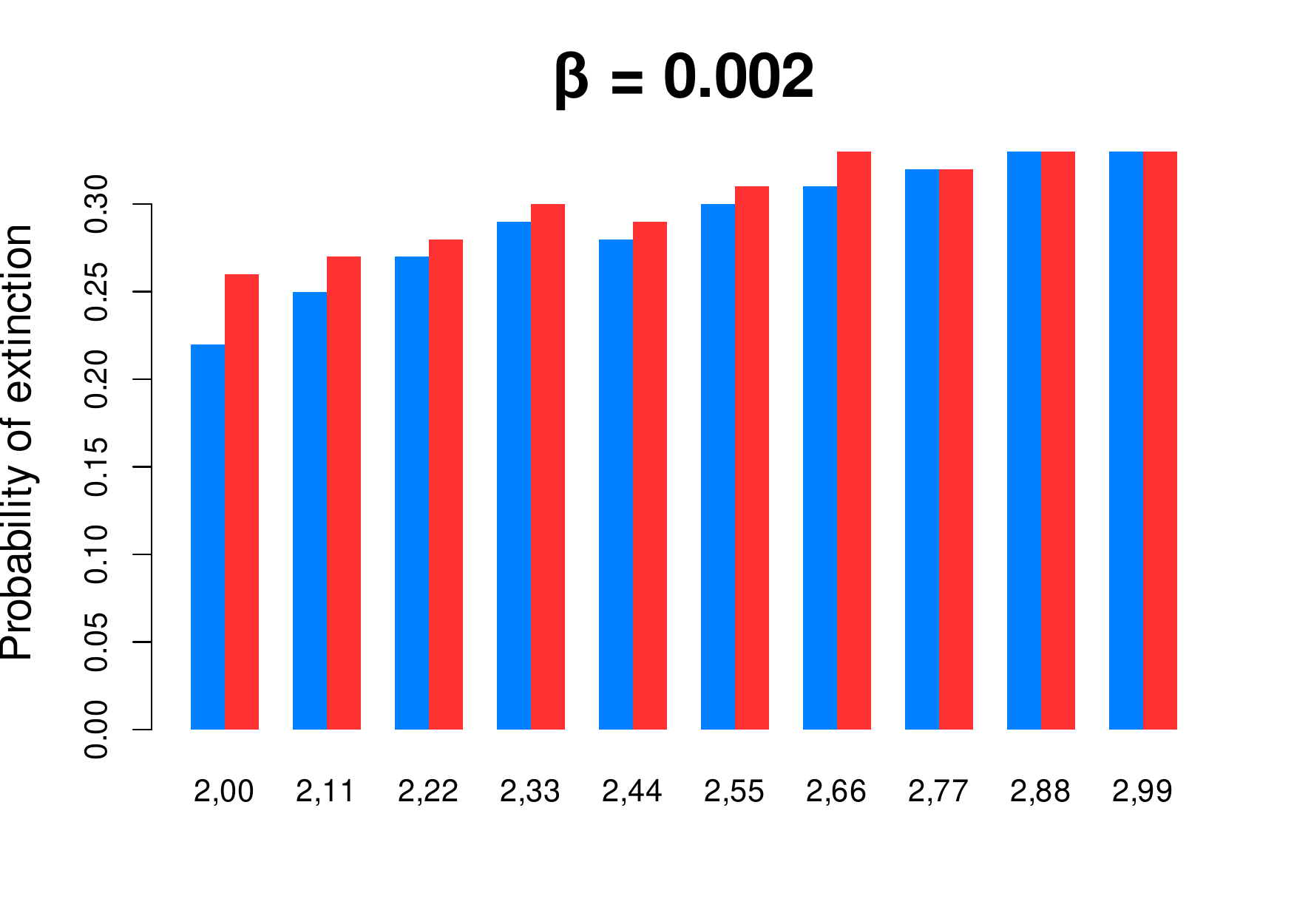}
	\end{subfigure}
	\hfill
	\begin{subfigure}[b]{0.49\textwidth}
		\centering
		\includegraphics[width=1\textwidth]{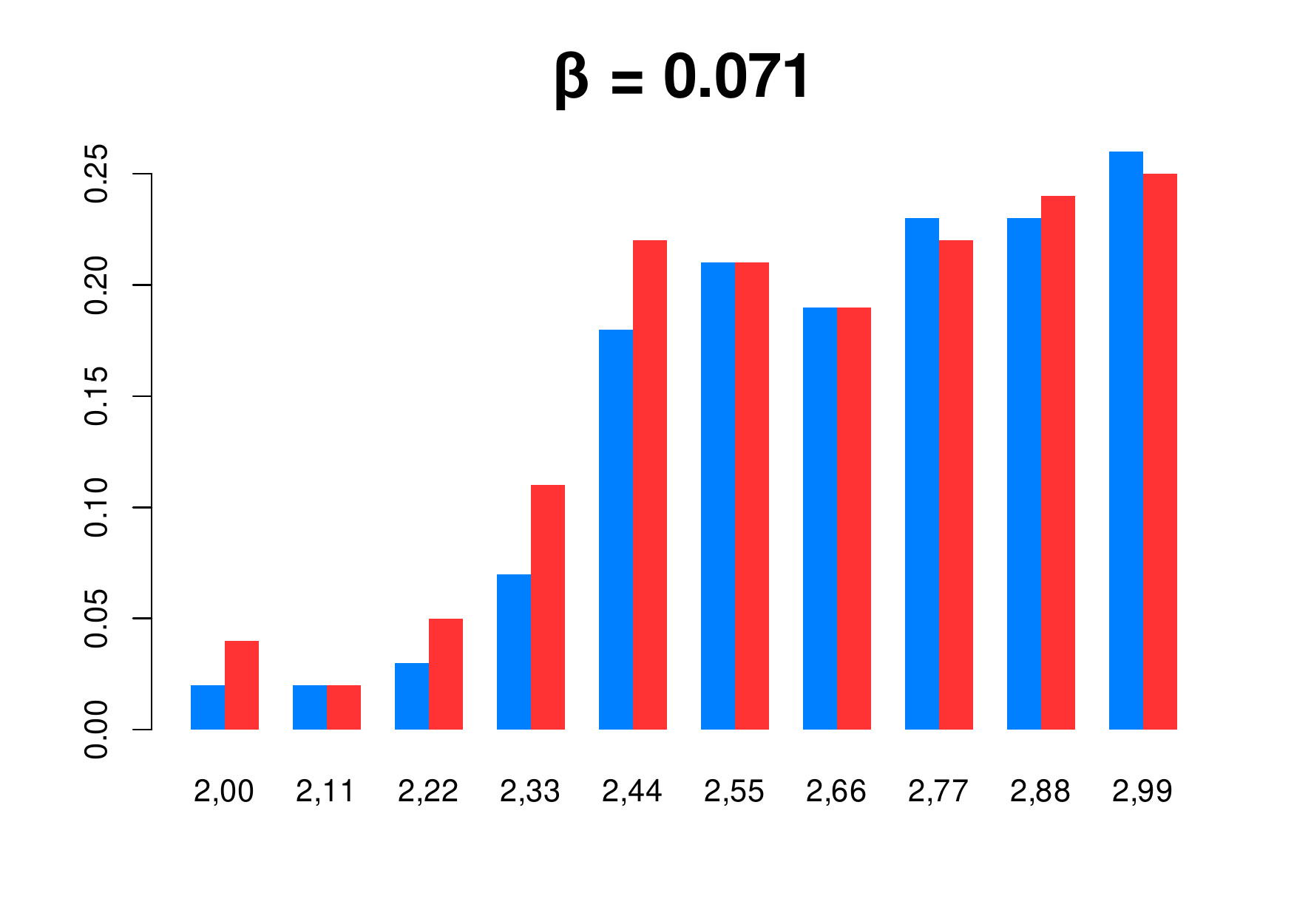}
	\end{subfigure}
\end{figure}
\begin{figure}[H]\ContinuedFloat
	\centering
	\begin{subfigure}[b]{0.49\textwidth}
		\centering
		\includegraphics[width=1\textwidth]{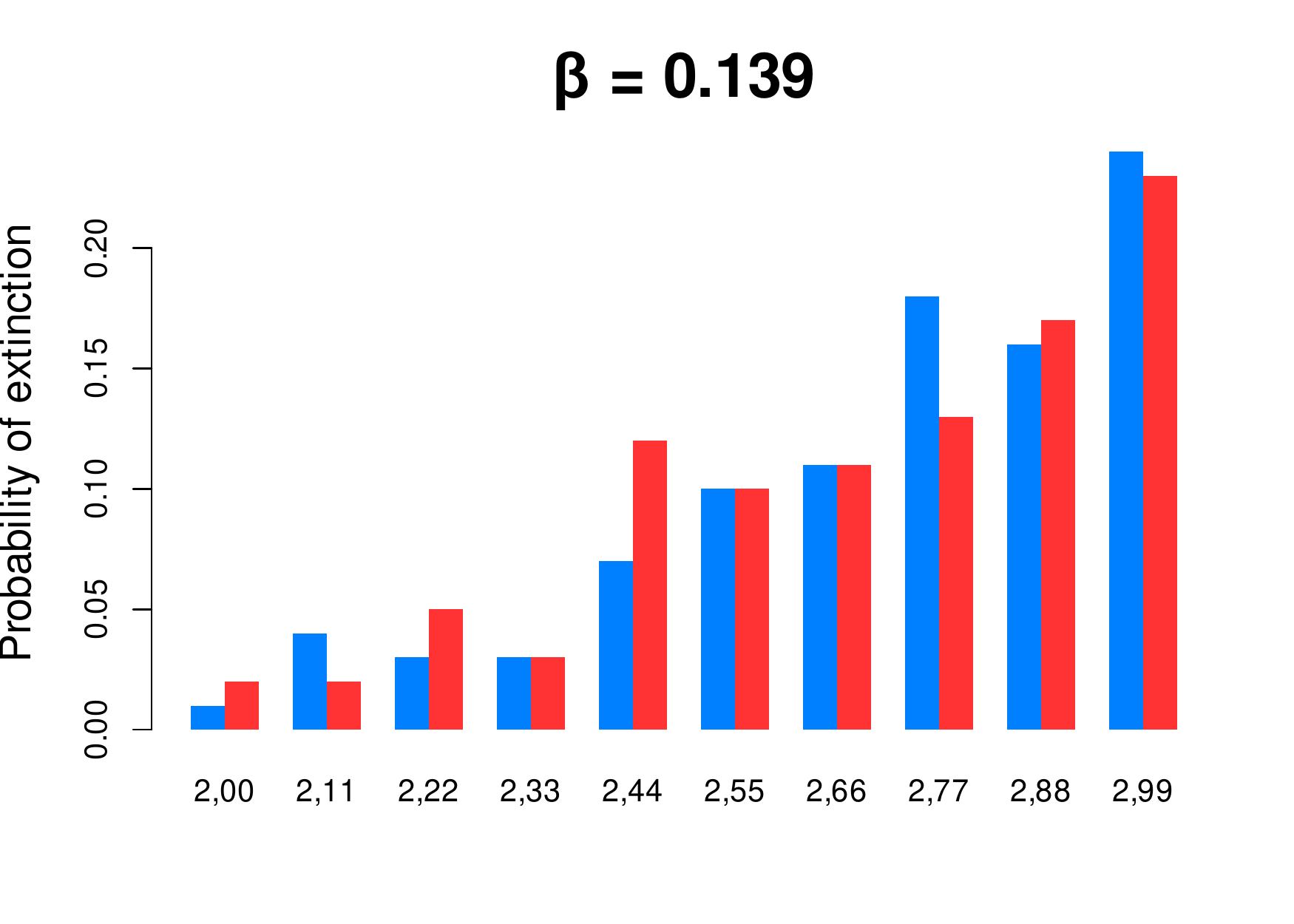}
	\end{subfigure}
	\hfill
	\begin{subfigure}[b]{0.49\textwidth}
		\centering
		\includegraphics[width=1\textwidth]{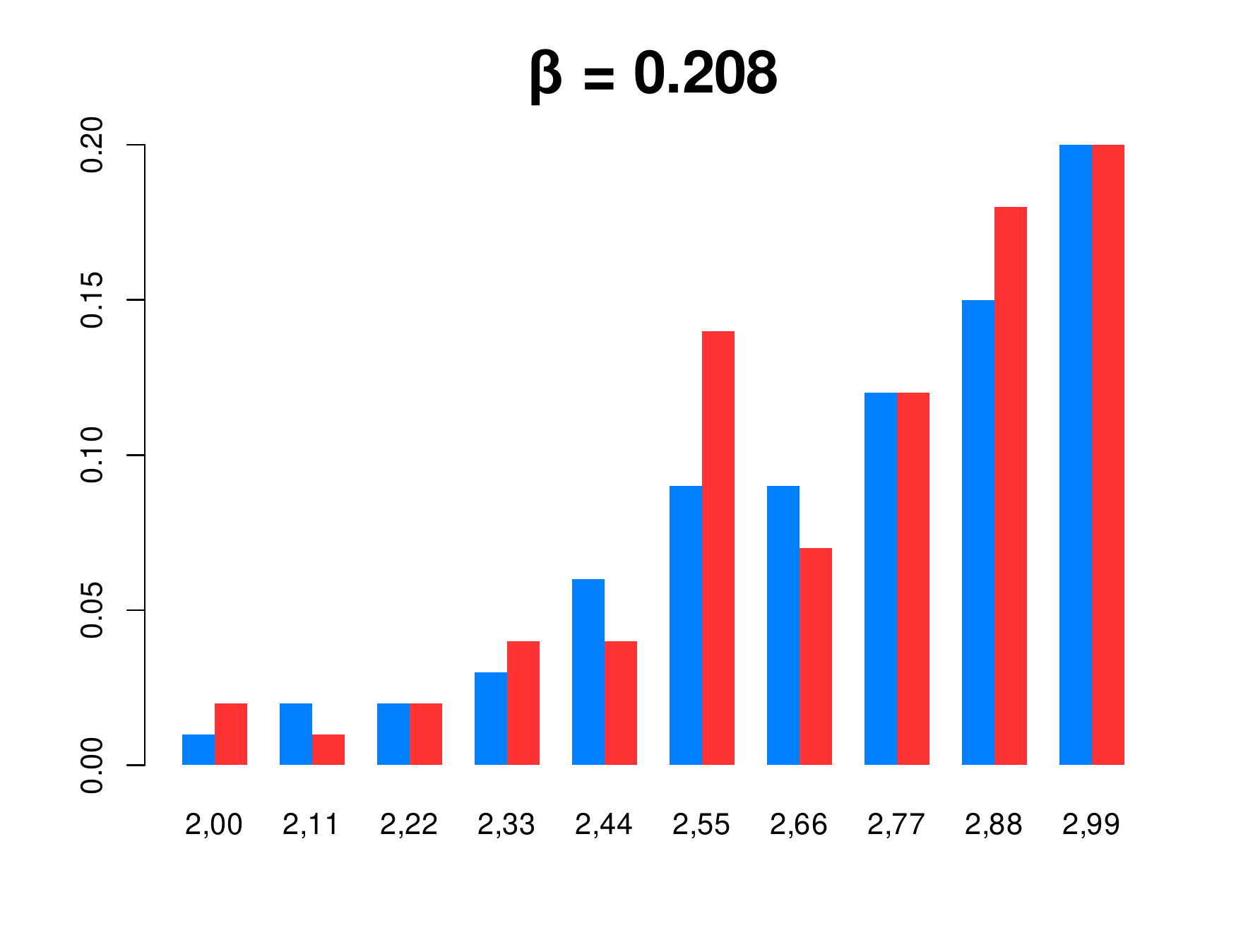}
	\end{subfigure}
\end{figure}
\begin{figure}[H]\ContinuedFloat
	\begin{subfigure}[b]{0.49\textwidth}
		\centering
		\includegraphics[width=1\textwidth]{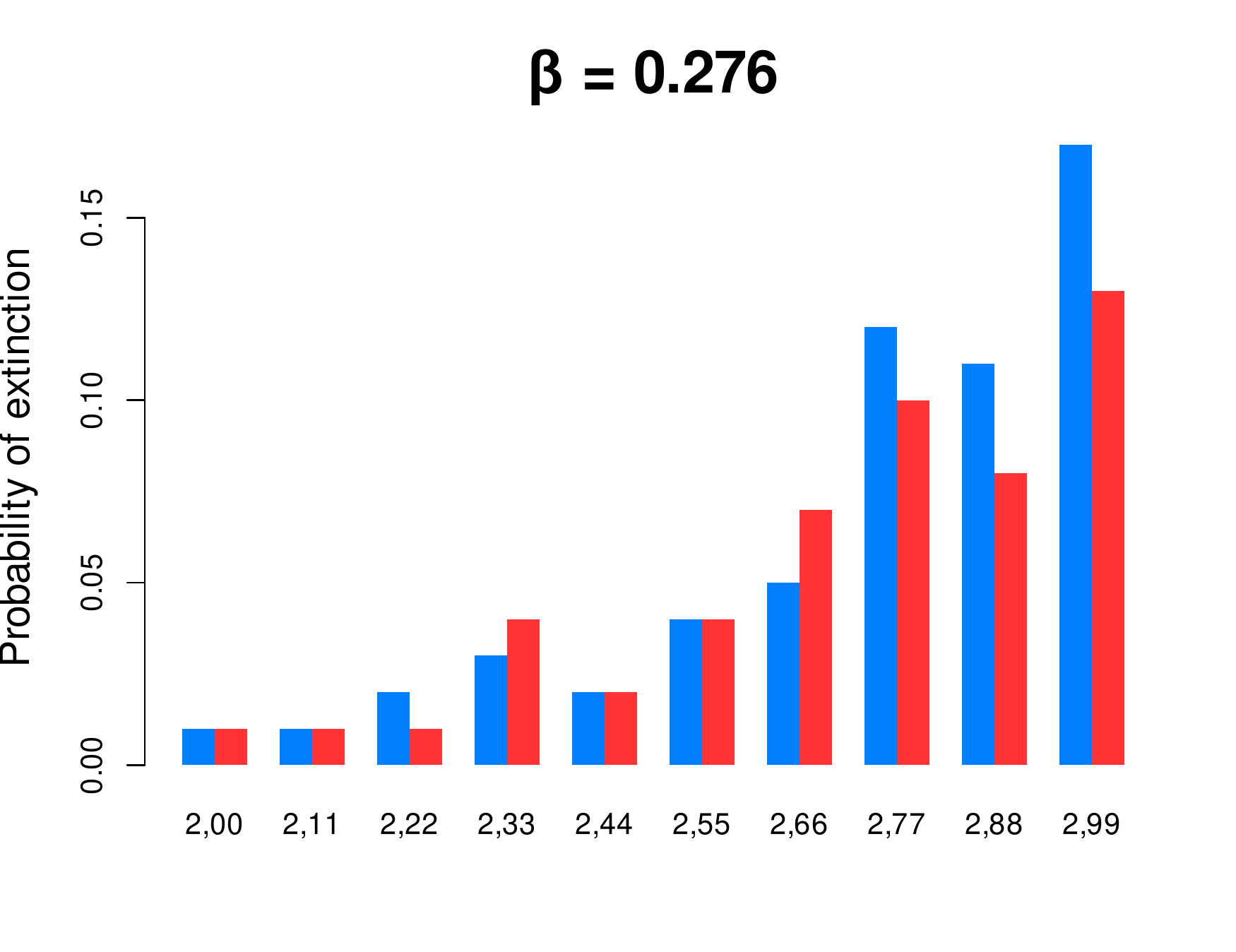}
	\end{subfigure}
	\hfill
	\begin{subfigure}[b]{0.49\textwidth}
		\centering
		\includegraphics[width=1\textwidth]{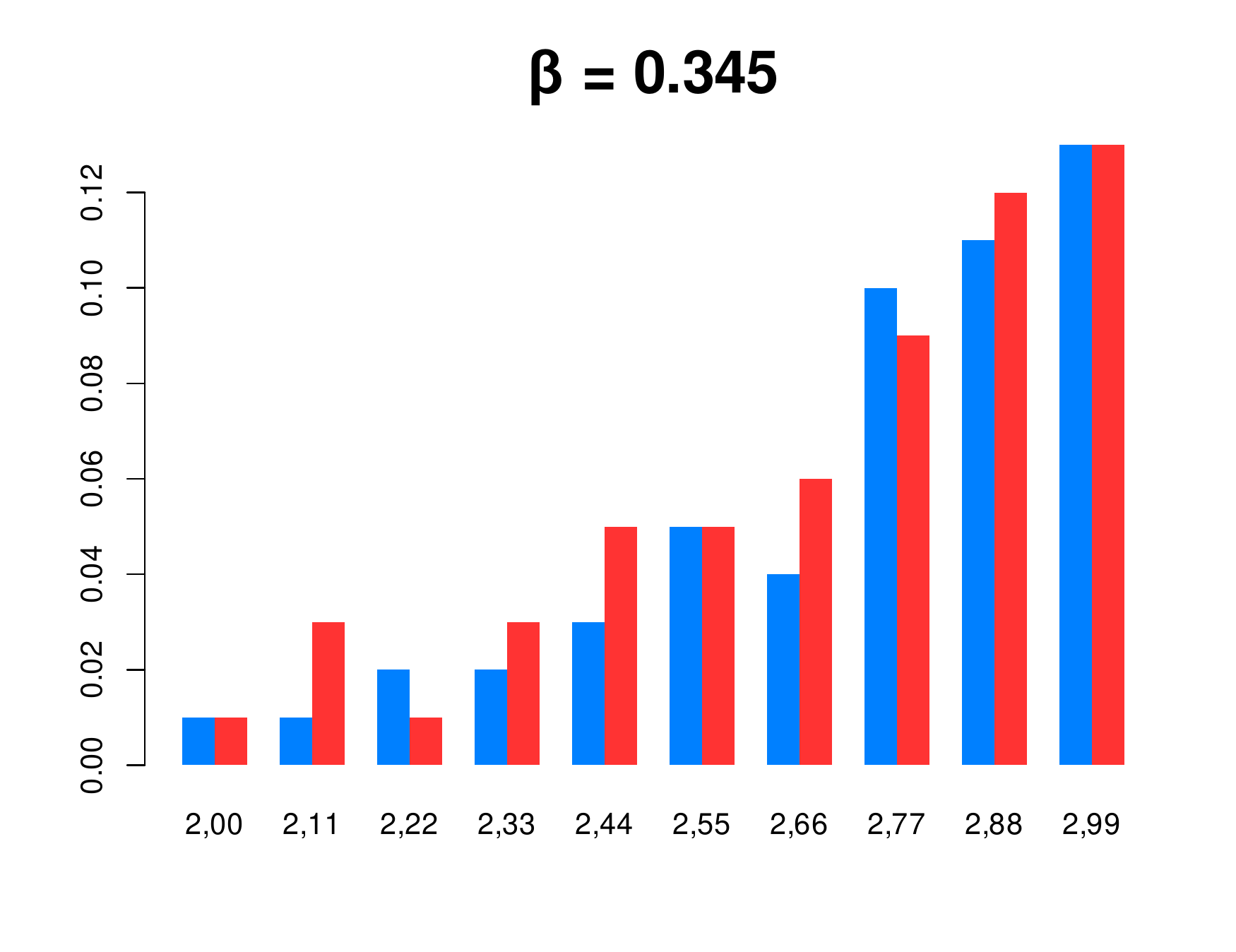}
	\end{subfigure}
\end{figure}
\begin{figure}[H]\ContinuedFloat
	\begin{subfigure}[b]{0.49\textwidth}
		\centering
		\includegraphics[width=1\textwidth]{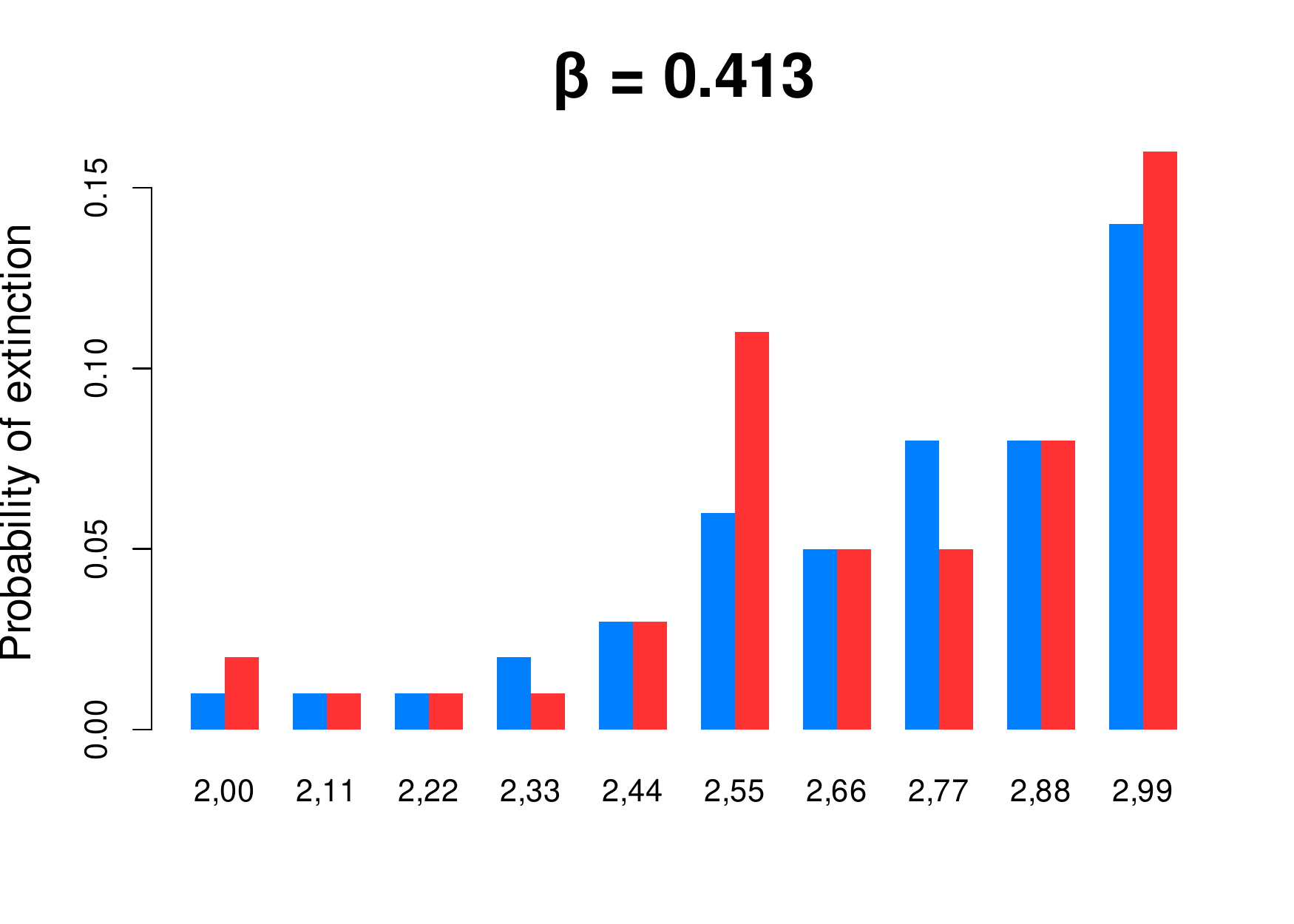}
	\end{subfigure}
	\hfill
	\begin{subfigure}[b]{0.49\textwidth}
		\centering
		\includegraphics[width=1\textwidth]{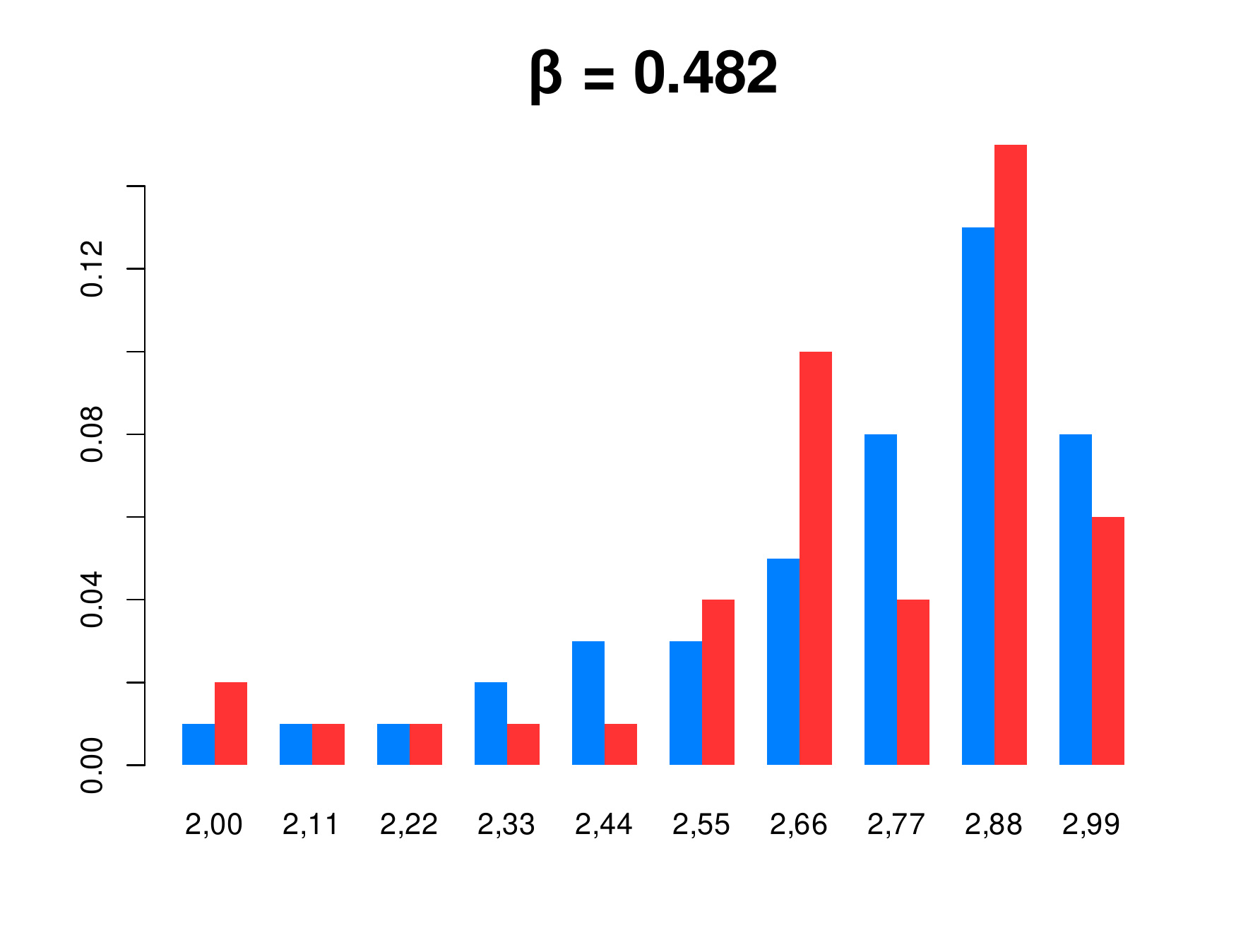}
	\end{subfigure}
\end{figure}
\begin{figure}[H]\ContinuedFloat
	\begin{subfigure}[b]{0.49\textwidth}
		\centering
		\includegraphics[width=1\textwidth]{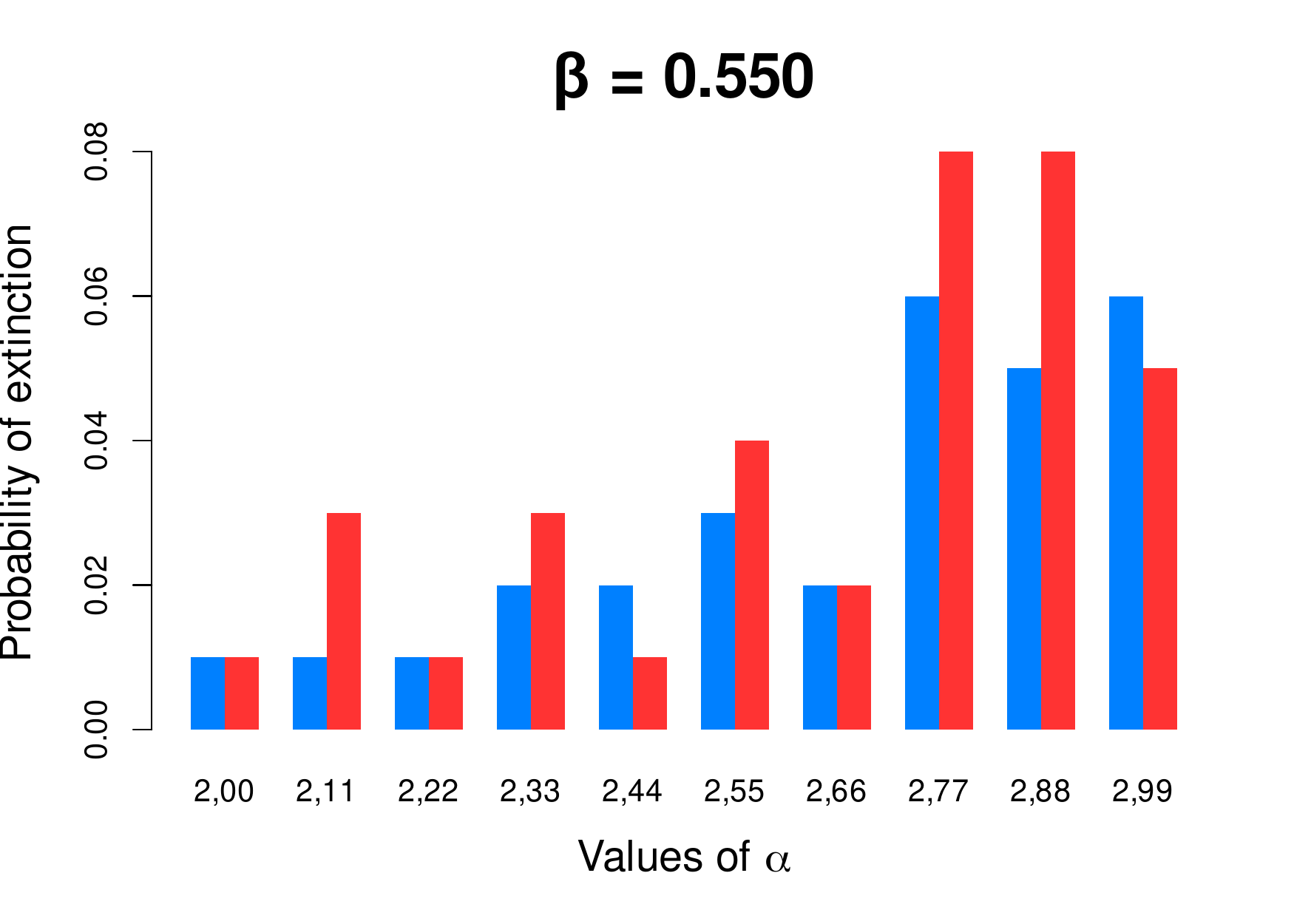}
	\end{subfigure}
	\hfill
	\begin{subfigure}[b]{0.49\textwidth}
		\centering
		\includegraphics[width=1\textwidth]{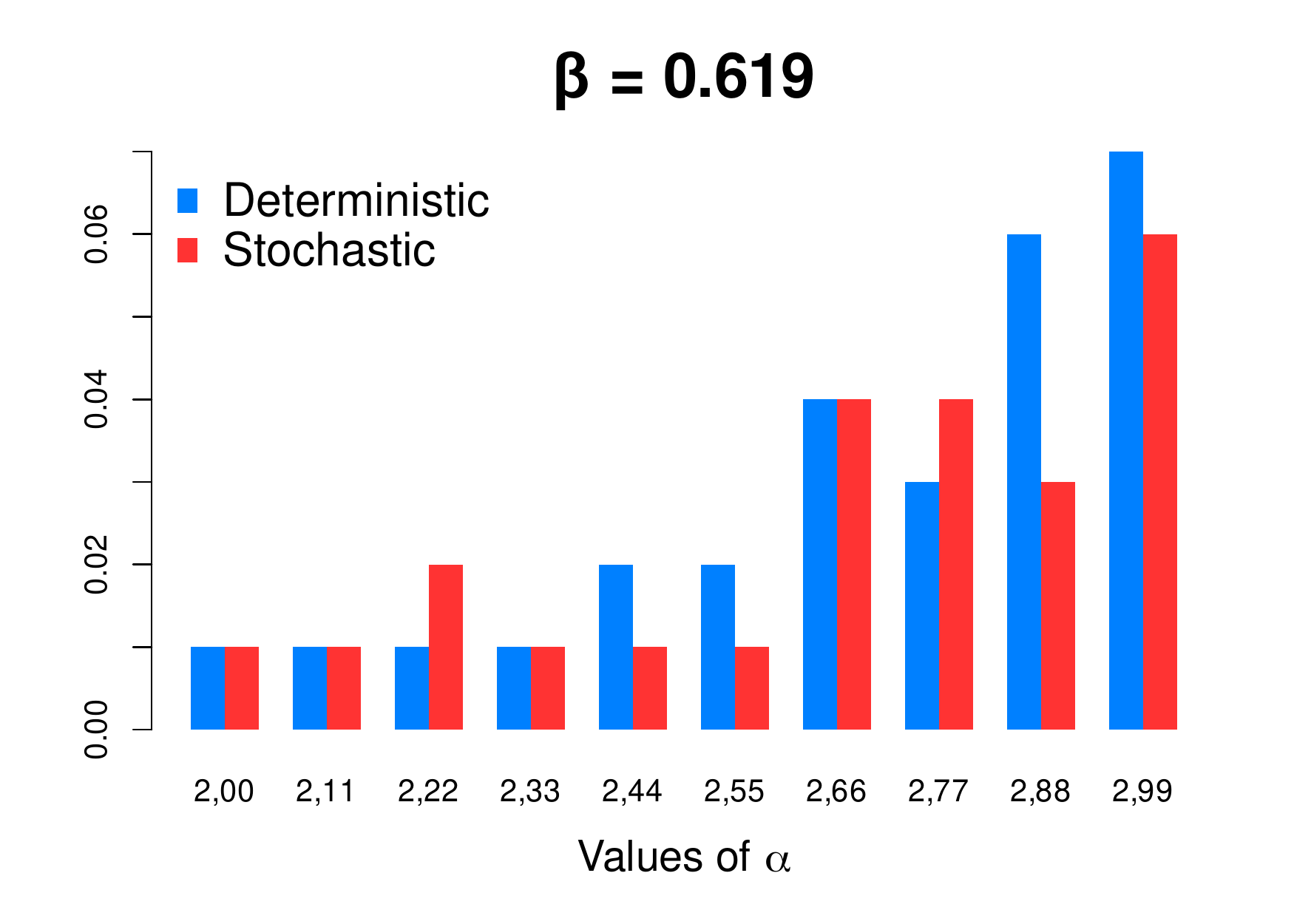} 
	\end{subfigure}
	\caption{Comparison between stochastic (red bars) and deterministic (blue bars) results. }
	\label{fig:comparison}
\end{figure}

\subsection{Different numbers of initially infected nodes}

\begin{figure}[H]
	\centering
	\begin{subfigure}[b]{1\textwidth}
		\centering
		\includegraphics[width=\textwidth]{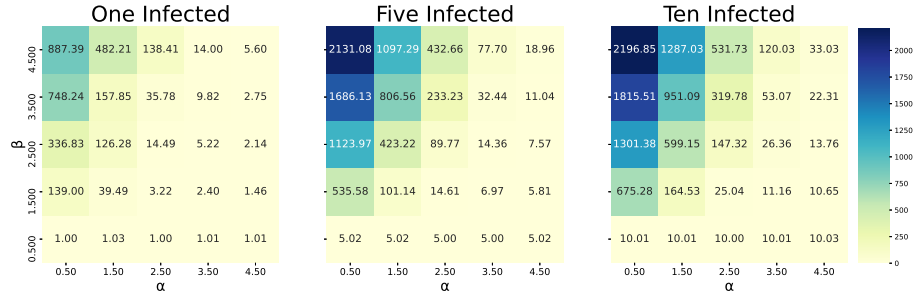}
		\caption{Maximum number of simultaneously infected nodes.}
		\label{fig:randcompamax}
	\end{subfigure}
	\hfill
	\begin{subfigure}[b]{1\textwidth}
		\centering
		\includegraphics[width=\textwidth]{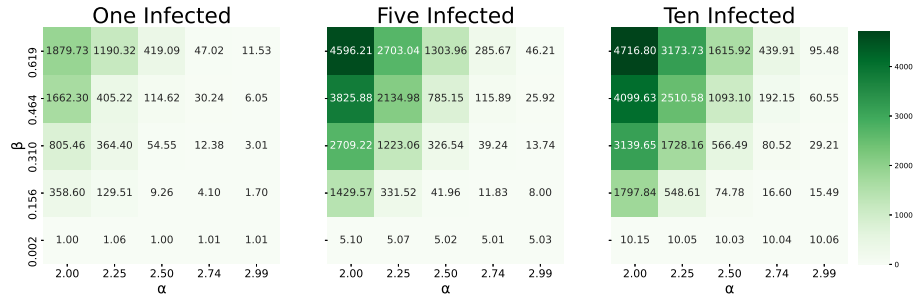}
		\caption{Number of eventually infected nodes.}
		\label{fig:randcompaeve}
	\end{subfigure}
	\hfill
	\begin{subfigure}[b]{1\textwidth}
		\centering
		\label{fig:comparison2}
		\includegraphics[width=\textwidth]{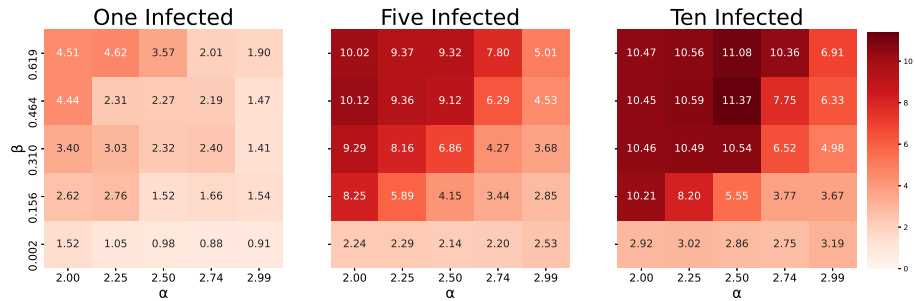}
		\caption{End time for the epidemic.}
		\label{fig:randcompatime}
	\end{subfigure}
	\caption{Random choice of initially infected.}
	\label{random}
\end{figure}

\begin{figure}[H]
	\centering
	\begin{subfigure}[b]{1\textwidth}
		\centering
		\includegraphics[width=\textwidth]{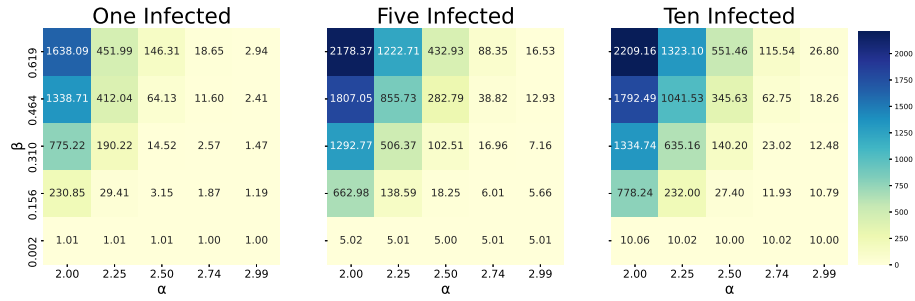}
		\caption{Maximum number of simultaneously infected nodes.}
		\label{fig:meancompamax}
	\end{subfigure}
	\hfill
	\begin{subfigure}[b]{1\textwidth}
		\centering
		\includegraphics[width=\textwidth]{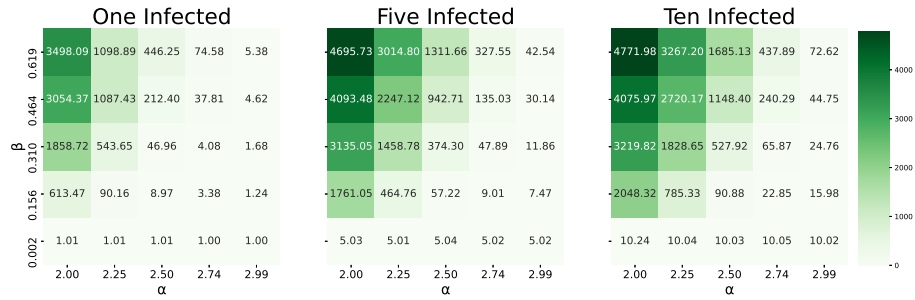}
		\caption{Number of eventually infected nodes.}
		\label{fig:meancompaeve}
	\end{subfigure}
	\hfill
	\begin{subfigure}[b]{1\textwidth}
		\centering
		\includegraphics[width=\textwidth]{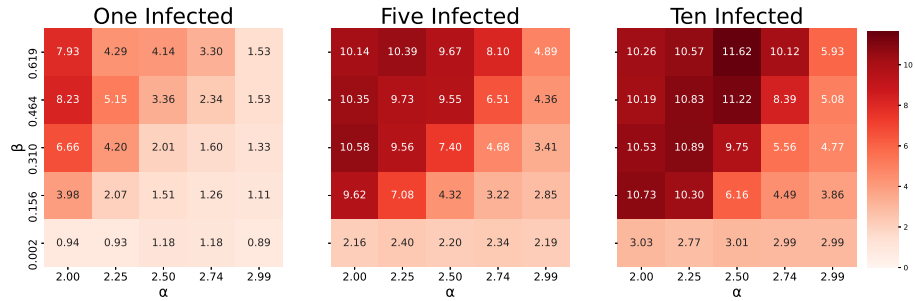}
		\caption{End time for the epidemic.}
		\label{fig:meancompatime}
	\end{subfigure}
	\caption{Mean-degree choice of initially infected.}
	\label{mean}
\end{figure}

\begin{figure}[H]
	\centering
	\begin{subfigure}[b]{1\textwidth}
		\centering
		\includegraphics[width=\textwidth]{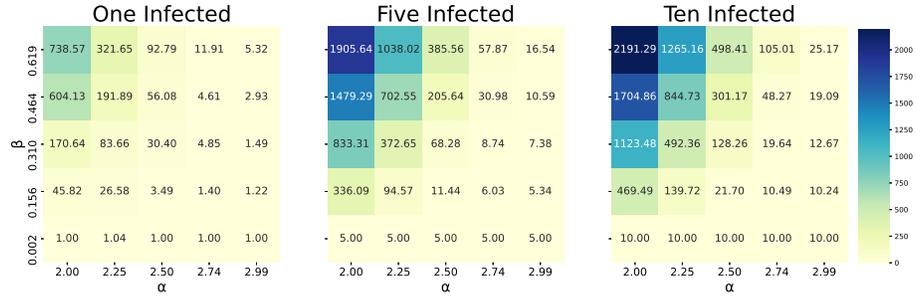}
		\caption{Maximum number of simultaneously infected nodes.}
		\label{fig:pericompamax}
	\end{subfigure}
	\hfill
	\begin{subfigure}[b]{1\textwidth}
		\centering
		\includegraphics[width=\textwidth]{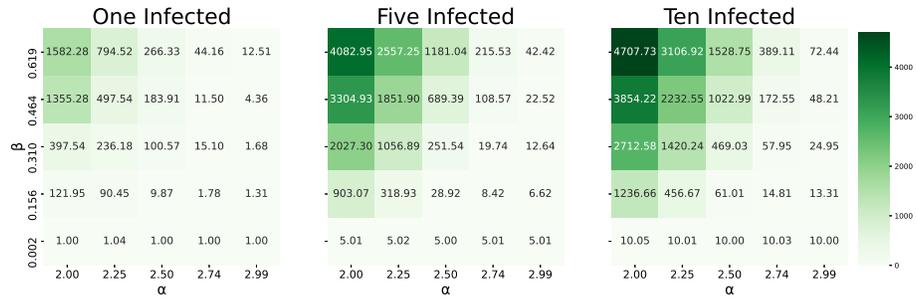}
		\caption{Number of eventually infected nodes.}
		\label{fig:pericompaeve}
	\end{subfigure}
	\hfill
	\begin{subfigure}[b]{1\textwidth}
		\centering
		\includegraphics[width=\textwidth]{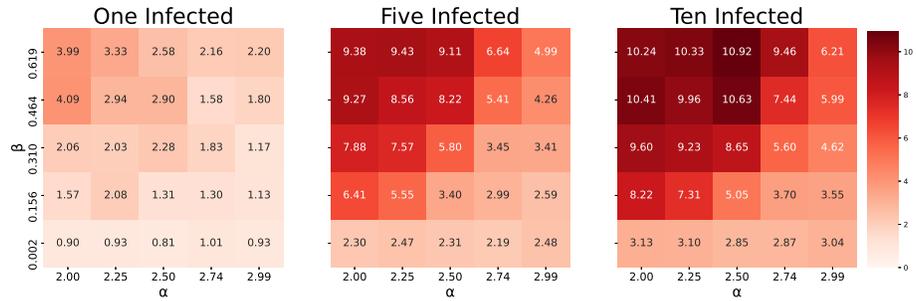}
		\caption{End time for the epidemic.}
		\label{fig:pericompatime}
	\end{subfigure}
	\caption{Peripheral-degree choice of initially infected.}
	\label{periph}
\end{figure}

\begin{figure}[H]
	\centering
	\begin{subfigure}[b]{1\textwidth}
		\centering
		\includegraphics[width=\textwidth]{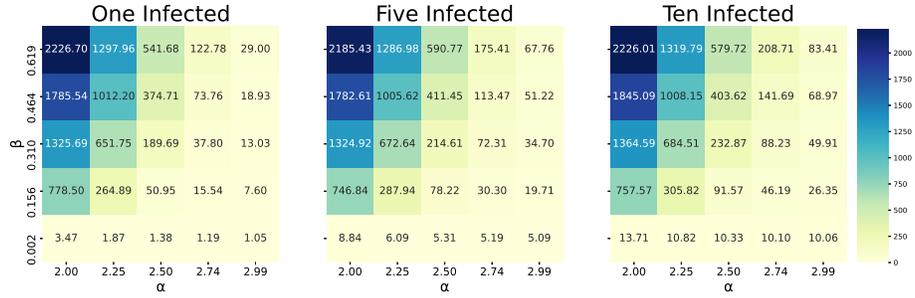}
		\caption{Maximum number of simultaneously infected nodes.}
		\label{fig:hubcompamax}
	\end{subfigure}
	\hfill
	\begin{subfigure}[b]{1\textwidth}
		\centering
		\includegraphics[width=\textwidth]{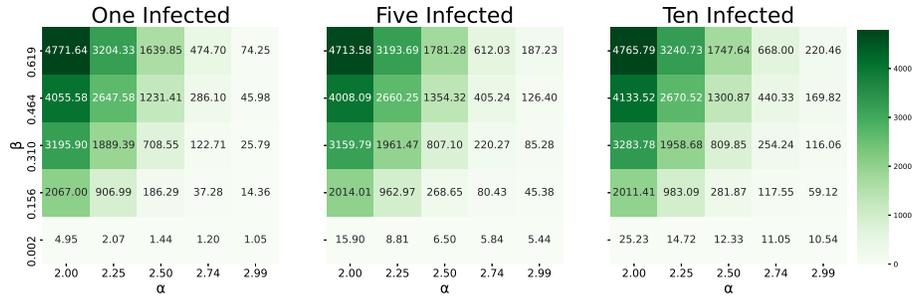}
		\caption{Number of eventually infected nodes.}
		\label{fig:hubcompaeve}
	\end{subfigure}
	\hfill
	\begin{subfigure}[b]{1\textwidth}
		\centering
		\includegraphics[width=\textwidth]{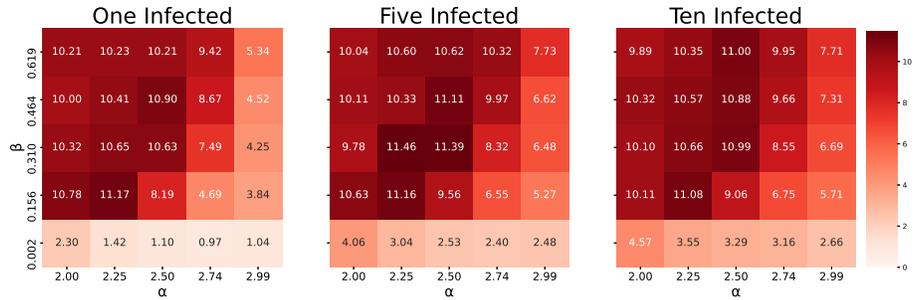}
		\caption{End time for the epidemic.}
		\label{fig:hubcompatime}
	\end{subfigure}
	\caption{Hub-degree choice of initially infected.}
	\label{hub}
\end{figure}

First, we considered as separate cases, epidemic starting with three different values of initially infected nodes, that are $I(0)=1$, $I(0)=5$ and $I(0)=10$, with four possible initial positions in the network: random, hub, intermediate, and peripheral; we compare the results in the heatmaps in Figures \ref{random}, \ref{mean}, \ref{periph} and \ref{hub}. 

If the position of the initially infected node changes, the dynamics of the spread of the disease mutates accordingly. Indeed, when the first infected nodes are on the periphery, the epidemic is diffusing slowly compared to the case in which the epidemic starts in a node with more contacts. The random case is qualitatively intermediate between the peripheral case and the mean-degree case: this follows from the distribution of the degree of nodes in the network, since we have few nodes in the tail of the distribution (hub).

The resulting numbers in the heatmaps do not seem to change significantly between the $I(0)=5$ and $I(0)=10$ case, suggesting a saturation effect already for a small (compared to the total population) number of initially infected nodes. Moreover, in the random, mean-degree and peripheral cases the number of infected (both simultaneously and eventually) is increasing as the initial number of infected nodes grows. 

The hub case represents a special case, as shown in Figure \ref{hub}. There, the number of the infected nodes is greater in the case of only one initially infected node than the other two cases. This behaviour can be explained in two distinct ways: first of all, the position of the first infected nodes are assigned in a ``hierarchical manner'', that is, since the hubs live in the tail of the power-law distribution, the second infected node inserted in the network has a degree lower than the first one. Moreover, if two neighbouring nodes are surrounded by several infected at the same time, they will be infected and the will recover approximately simultaneously, creating a certain saturation in the number of available susceptible nodes. 

The hub case is a ``worst case scenario'': the epidemic spreads even if it starts from only one infected node, if it has enough strength in terms of its parameters to do so. Moreover, it is very unlikely that more than one infected node enters a susceptible population \emph{simultaneously}. Hence, in the next section, we analyze the hub scenario with $I(0)=1$ in greater detail.

As expected, the epidemic spreads more with an increase in $\beta$, and less with an increase in $\alpha$, recalling that a greater $\alpha$ indicates an overall less connected network. Moreover, the more central the initial infection is, the more the disease is able to spread through the nodes of the network.

We notice a noisiness of the final time, compared to the other quantities, which is known in literature and to be expected, since the final part of the epidemic is a subcritical branching process, which is characterized by a great variability in its duration.

We notice that the average of maximum simultaneously and eventually infected nodes is increasing in $\beta$, and decreasing in $\alpha$; this trend becomes more evident the higher the number of initially infected nodes. This is due to the fact that the higher number of initially infected nodes, the smaller the probability of \textit{all of them} being in a small hubs, disconnected from the giant cluster.

\subsection{Initially infected nodes $I(0)=1$}

Figures \ref{fig:het1,2} to \ref{fig:het2,2} reflect what we expect to happen during an epidemic on a network. Higher values of $\beta$ correspond to higher values of $R_0$ (recall (\ref{eqn:rnod})), thus causing a greater spread of the epidemic for fixed values of $\alpha$: this is evident by looking at columns of the following heatmaps. The opposite is true for $\alpha$: greater values of $\alpha$ mean a less connected network, which impairs the spread of the disease as it can be noticed by looking at rows of the heatmaps; it is less evident, from equation (\ref{eqn:rnod}), at least, that greater values of $\alpha$ imply smaller values of $R_0$, but our simulations confirm the intuition and what was known in the literature for this model \cite{Barabasi_book}.

Clearly, the more central and connected the initially infected node is, the greater the magnitude of the infection, as the heatmaps labelled ``hub'', ``mean degree'' and ``peripheral'' illustrate; the ``random'' heatmaps give, qualitatively, an average of the other three.

The only quantity which remains noisy is the final time, visualized in \ref{fig:het3,2}: the tendencies are the same as in the two previous figure, i.e. values decreasing with $\alpha$ and increasing with $\beta$, but the resulting heatmap is quite noisy.

\begin{figure}[H]
	\centering
	\includegraphics[width=\textwidth]{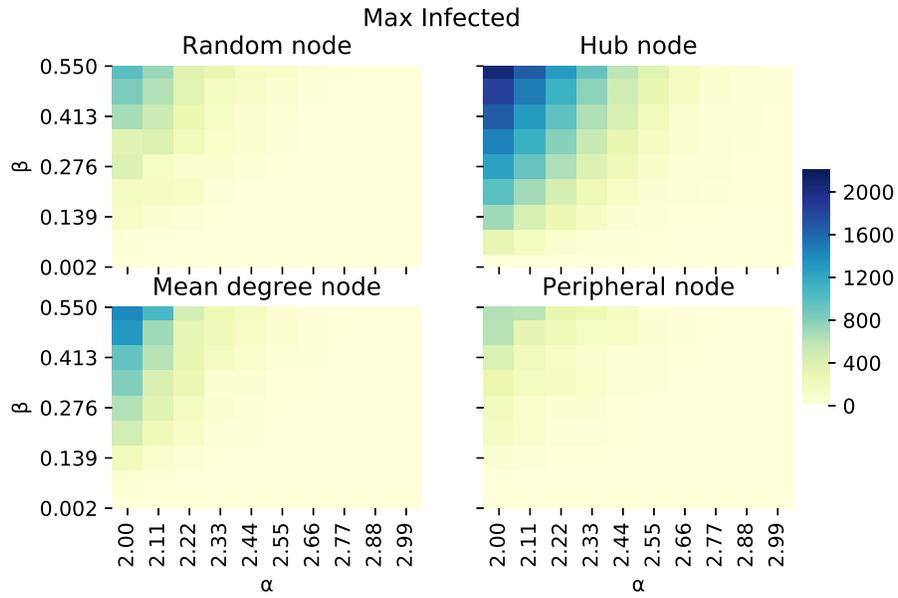} 
	\caption{Maximum number of simultaneously infected nodes.}
	\label{fig:het1,2}
\end{figure}

\begin{figure}[H]\centering
	\includegraphics[width=\textwidth]{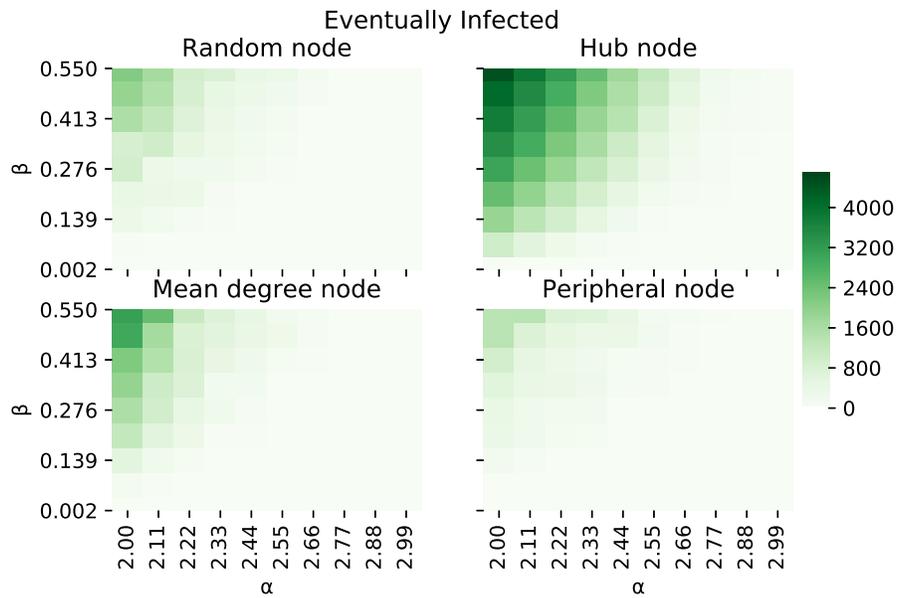}
	\caption{Number of eventually infected nodes.}
	\label{fig:het2,2}
\end{figure}

\begin{figure}[H]\centering
	\includegraphics[width=\textwidth]{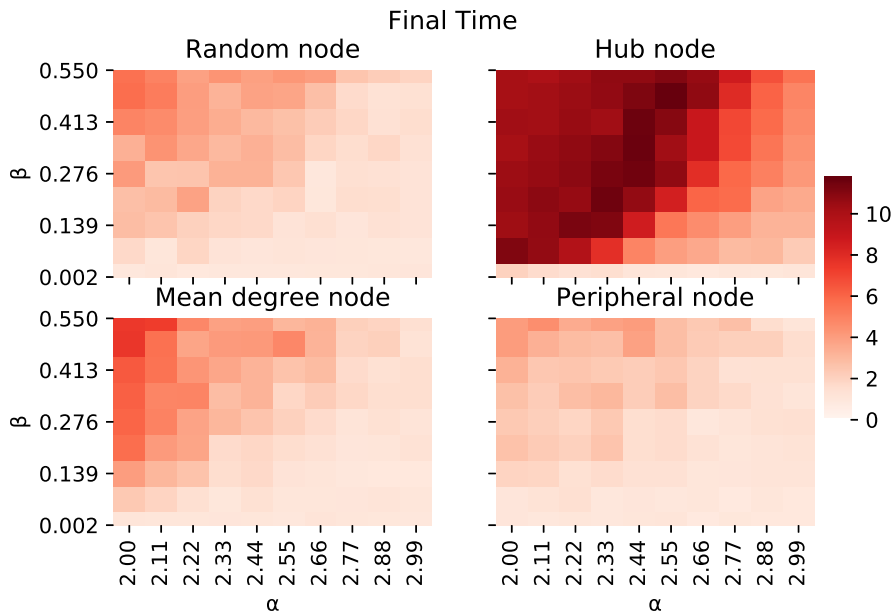}
	\caption{End time for the epidemic.}
	\label{fig:het3,2}
\end{figure}

\section{Discussion}\label{Concl}
We have shown how an epidemic effects a synthetic network, following a cut-off power law distribution for the degree of the nodes through the use of the configuration model. 
We provided a deterministic result on the probability of extinction of the infectious disease, which is based on the initial condition of the network and on the initially infected node. Our deterministic results and stochastic simulation results are in good agreement; the gap in the values stems from the fact that we have considered an infinite network in the deterministic framework, and a finite network, yet with a large number of nodes, in the stochastic one.

We have analysed various possible initial conditions for the networks we simulated via heatmaps, and compared them with the available deterministic insight. In particular, we explored how the position of the initially infected individual influences the whole epidemic, measured through three indices: eventually infected individuals, maximum simultaneously infected individuals and overall duration of the epidemic. 
Our analysis and numerical exploration confirm that the infectiousness of the disease is directly proportional to the spread of the epidemic; moreover, we conclude that the same disease (i.e., characterized by the same parameters $\beta$ and $\gamma$) infects more individuals in networks generated with a lower exponent power-law, as one would expect intuitively.

We have simulated the epidemic as a SIR stochastic dynamic using a specialised version of the Gillespie algorithm.  Our algorithm is versatile; many different additional features can be implemented, for example, dynamics edges, contact tracing, quarantine.  We leave these as inspiration for future work.

\section*{Acknowledgements}
The authors thank Prof. Andrea Pugliese for his helpful comments and 
expert help in various stages of the manuscript.

\bibliographystyle{elsarticle-num-names} 
\bibliography{biblio}

\end{document}